%% file: Main.tex
\documentclass[a4paper,11pt]{article}

\input{header.tex}

\title{Distribution of the transfer matrix in disordered wires}

\author{Mark Ancliff}

\begin{document}

\bibliographystyle{unsrt}
\maketitle

\begin{quote}

A closed expression is derived for the probability distribution of the transfer matrix of a particle moving in a one-dimensional system with delta-correlated, weak disorder. The change in the distribution as a function of wire length is described by a diffusion equation on the $SU(1,1)$ group, which is solved through the decomposition of the regular representation into irreducible components. The expression generalizes a number of well-known results, including the distributions of the transmission coefficient and local density of states. As an application, the average single energy-level contribution to the persistent current in a flux-threaded ring is derived.

\end{quote}

\section{Introduction}

The one-dimensional disordered wire represents the simplest example of a disordered mesoscopic system. Over the past fifty years it has been analyzed using a variety of techniques, including diagrammatics\cite{Berezinskii1974,Gogolin1982,Altshuler1989}, Fokker-Planck equations\cite{Abrikosov1981}, and functional-integration\cite{Kolokolov1994}. The system is unusual in that exact expressions can be obtained for many quantities, such as distribution of the transmission coefficient\cite{Abrikosov1981}, distribution of local density of states\cite{Altshuler1989}, and density-density correlators\cite{Gogolin1982}.

The system can be described through the transfer matrix, $T_E(x,x')$, which defines the transport of electrons at energy $E$. The main result of this paper is a derivation of the distribution of the transfer matrix $T_E$ in the case of short-range, Gaussian disorder. This distribution entails known results for the transmission coefficients, local density of states, and density-density correlators (at a single energy), but is not entailed by them, and as such represents a new exact result for the disordered wire.

The result is obtained by utilizing the representation theory of the pseudo-unitary group $SU(1,1)$. Conservation of the probability current implies that the transfer matrix must belong to this group, and therefore the distribution of transfer matrices belongs to the regular representation of functions on the group. All calculations of physical quantities at a single energy can be rewritten as problems in the representation theory. 

The paper is organized as follows. In section \ref{SecDiffusion} we derive a diffusion equation on the group for the distribution of transfer matrices as a function of wire length. In section \ref{SecProbDist} we solve the diffusion equation by decomposing the regular representation into irreducible unitary subrepresentations. Section \ref{SecProperties} analyses the distribution and shows how the distributions of other physical properties can be obtained from it. Finally, as an application, section \ref{SecRing} considers the case of a disordered ring. Specifically, we calculate the average contribution to the persistent current from electrons in a given energy interval. The paper concludes with a discussion of possible generalizations of the representation-theoretic method presented here to other systems. 

\section{Diffusion equation for the probability distribution}\label{SecDiffusion}

We choose units so that $\hbar = 2m = 1$ and consider the one-dimensional Hamiltonian, $H = - \pdd{}{x^2}{} + V(x)$, where $V(x)$ is Gaussian distributed and $\expected{V(x)V(x')} = \alpha^2 \delta(x-x')$. For weak disorder it is reasonable to write an Eigenstate of energy $E$ as a sum of right- and left-moving parts,
\begin{equation}
\psi(x) = \phi_R(x) \ue^{\ui k x} + \phi_L(x) \ue^{-\ui k x}
\end{equation}
where $k = \sqrt{E}$, and $\phi_R(x)$ and $\phi_L(x)$ are slowly changing functions (i.e. $|\phi_{R/L}'(x)/\phi_{R/L}(x)| \ll k$). The equation $H\psi = E\psi$ is equivalent to the condition
\begin{equation}\label{EigenEquation}
\sd{}{x} \begin{pmatrix} \phi_R \\ \phi_L \end{pmatrix} = - \frac{V(x)}{2 k} \begin{pmatrix} \ui & \ui \ue^{-\ui 2 k x} \\ -\ui \ue^{\ui 2 k x} & -\ui \end{pmatrix} \begin{pmatrix} \phi_R \\ \phi_L \end{pmatrix}
\end{equation}
In what follows we denote the matrix occuring on the right-hand side of (\ref{EigenEquation}) by $\hat{A}(x)$. The transfer matrix $\hat{T}(x,x')$ between any two points $x$ and $x'$ on the wire is defined by the condition $\vec{\phi}(x) = \hat{T}(x,x')\vec{\phi}(x')$. From equation (\ref{EigenEquation}) it follows that
\begin{equation}\label{TransferMatrix}
\sd{}{x}\hat{T}(x,x') = -\frac{V(x)}{2k} \hat{A}(x)\  \hat{T}(x,x')
\end{equation}
It can be checked that the matrix $\hat{A}(x)$ lies in the $su(1,1)$ algebra, and therefore (\ref{TransferMatrix}) is a differential equation on the $SU(1,1)$ group.

Equation (\ref{TransferMatrix}) is a significant generalization of equation (\ref{EigenEquation}): while in the former the representation on which the group acts is fixed (as the two-dimensional `natural' representation), the latter refers only to the group itself, and we are therefore free to consider its action on an arbitrary state of an arbitrary representation.

In particular, we consider the (left-) regular representation and let $\ket{\delta}$ denote the delta-function\footnote{Strictly speaking the delta `function' is a functional and does not belong to the regular representation of $L^2$-normalizable functions. A more rigorous presentation would involve the construction of a rigged Hilbert space.} at the identity with respect to the Haar measure of $SU(1,1)$. The action of $\hat{T}(x,x')$ shifts the delta-function from the identity to the point $\hat{T}(x,x')$. Therefore, given an ensemble of transfer matrices, the ensemble average $\expected{\hat{T}(x,x')}$ applied to the state $\ket{\delta}$ gives the probability distribution of transfer matrices with respect to the Haar measure, i.e.
\begin{equation}\label{ProbDist}
P\left( \hat{T},x-x'\right) = \left( \expected{\hat{T}(x,x')}\ket{\delta}\right)(\hat{T})
\end{equation}

It remains to relate the ensemble of transfer matrices to the statistics of the disordered potential $V(x)$. Since the disorder is weak we can consider $x = x'+\delta x$ such that $\delta x \gg 1/k$ but short enough that $\hat{T}(x,x')\approx \hat{1}$. The Transfer matrix can be expanded as
\begin{equation}
\hat{T}(x,x') = \hat{1} - \frac{1}{2k}\int_{x'}^x V(s)\hat{A}(s) \ud s + \frac{1}{4k^2}  \int_{x'}^x \int_{x'}^s V(s)V(t)\hat{A}(s)\hat{A}(t) \ud s \ud t + \cdots
\end{equation}
Higher order terms do not contribute to first order in $\delta x$. Since $\expected{V} = 0$ and $\expected{V(s)V(t)} = \alpha^2 \delta(s-t)$, upon averaging over disorder we obtain
\begin{equation}
\expected{\hat{T}(x,x')}  =  \hat{1} + \frac{\alpha^2}{8k^2}\int_{x'}^x \hat{A}(s)^2 \ud s + O(\delta x^2) 
\end{equation}
Choosing a basis for the $su(1,1)$ algebra,
\begin{equation}
\hat{L}_x = \begin{pmatrix} 0 & 1 \\ 1 & 0 \end{pmatrix} , 
\hat{L}_y = \begin{pmatrix} 0 & -\ui \\ \ui & 0 \end{pmatrix} , 
\hat{L}_z = \begin{pmatrix} \ui & 0 \\ 0 & -\ui \end{pmatrix}
\end{equation}
and rewriting $\hat{A}(s)$ in this basis gives
\begin{eqnarray}
\expected{\hat{T}(x,x')} & = & \hat{1} + \frac{\alpha^2}{8k^2}\int_{x'}^x \big(\hat{L}_z + \sin(2ks) \hat{L}_x - \cos(2ks) \hat{L}_y \big)^2 \ud s + O(\delta x^2) \nonumber \\
& = & \hat{1} + \frac{\alpha^2}{16k^2} \big( 2{\hat{L}_z}^2 + {\hat{L}_x}^2 + {\hat{L}_y}^2 \big) \delta x + O(\delta x^2)
\end{eqnarray}
where in the last line we have ignored terms of order $1/k$. Thus, (for arbitrary $x$ and $x'$) we have
\begin{equation}\label{TavEvolution}
\sd{}{x}\expected{\hat{T}(x,x')} =  \frac{\alpha^2}{16k^2} \big( 2{\hat{L}_z}^2 + {\hat{L}_x}^2 + {\hat{L}_y}^2 \big) \expected{\hat{T}(x,x')}
\end{equation}
Combining (\ref{TavEvolution}) with (\ref{ProbDist}) gives
\begin{equation}\label{ProbDiffusion}
\sd{}{x}P\left( \hat{T},x-x'\right) = \frac{\alpha^2}{16k^2} \big( 2{\hat{L}_z}^2 + {\hat{L}_x}^2 + {\hat{L}_y}^2 \big) P\left( \hat{T}, x-x' \right)
\end{equation}
which should be solved with the initial condition $P\left( \hat{T},0 \right) = \delta(\hat{T})$. In the regular representation the algebra elements $\hat{L}_i$ describe first-order differential operators on the $SU(1,1)$ group, and interpreting $x$ as a time coordinate, (\ref{ProbDiffusion}) is identified as a diffusion equation for the probability. The parameter $\lambda = 4k^2/\alpha^2$ can be identified as the mean free path of the particle for back-scattering (in this case the rates of back- and forward- scattering are equal).

The solution to the diffusion equation given below relies heavily on the reprsentation theory of $SU(1,1)$, so the following section begins with a summary of the necessary results (for a thorough overview, see, e.g. \cite{Vilenkin1991}).

\section{Derivation of the probability distribution}\label{SecProbDist}

Given any group and a space of functions on the group, one can define two different group-actions on the space: action on the left, $\big(\hat{T}_1^{(L)} f \big) (\hat{T}_2) = f(\hat{T}_1^{-1} \hat{T}_2)$; and action on the right, $\big( \hat{T}_1^{(R)} f \big) (\hat{T}_2) = f(\hat{T}_2 \hat{T}_1)$. The regular representation can be decomposed as a sum of irreducible unitary representations with respect to either of these actions, and as the actions commute the right action maps between equivalent irreducible representations of the left action, and vice-versa.

Now let $V_\mu$ be a irreducible unitary representation. The Eigenstates of $\hat{L}_0 = -\frac{i}{2}\hat{L}_z$ form a basis for each $V_\mu$, and the fact that $\exp[2\pi \hat{L}_z] = \hat{1}$ implies that the Eigenvalues of $\hat{L}_0$ must be integer or half-integer (known as the \emph{weights} of the representation). Irreducibility implies that for each weight $n$ there is at most one Eigenstate of $\hat{L}_0$ in $V_\mu$. The notation $\ket{n^\mu}$ is used to represent the state of weight $n$ in the representation $V_\mu$, normalized so that $\braket{n^\mu}{n^\mu} = 1$.

The functions
\begin{equation}
t_\mu^{m,n}(\hat{T}) = \bra{m^\mu}\hat{T}\ket{n^\mu}
\end{equation}
span a representation of $SU(1,1)$ with respect to the left and right group actions given above. Specifically, they span a tensor product of a copy of $V_\mu$ with respect to the right action and a copy of the dual representation $V_\mu^\ast$ with respect to the left action, denoted $V_\mu^{\ast(L)}\otimes V_\mu^{(R)}$. Note that $t_\mu^{m,n}$ is a state with weight $-m$ with respect to the left action and weight $n$ with respect to the right action. The key to the decomposition of the regular representation is to find irreducible unitary representations $V_\mu$ such that
\begin{equation}
V_\text{regular} = \sum_\mu V_\mu^{\ast(L)}\otimes V_\mu^{(R)}
\end{equation}

The Casimir operator of the $su(1,1)$ algebra is given by
\begin{equation}
\hat{c} = \tfrac{1}{4}\left( \hat{L}_x^2 + \hat{L}_y^2 - \hat{L}_z^2 \right)
\end{equation}
and acts on a scalar on all irreducible representations. The irreducible representations occuring in the decomposition of the regular representation can be specified uniquely by the weights that occur within them and the value of the Casimir operator. They are found to be \cite{Vilenkin1991}:
\begin{enumerate}
\item The lowest-weight representations $V_{l,+}$ with weights $\{l+n | n \in \mathbb{Z}, n \geq 0 \}$, where $l$ is a positive (half-) integer, on which the Casimir operator takes the value $l(l+1)$;
\item The highest-weight representations $V_{l,-}$ with weights $\{-(l+n) | n \in \mathbb{Z}, n \geq 0 \}$, where $l$ is a positive (half-) integer, on which the Casimir operator takes the value $l(l+1)$ as above (these are the dual representations of the lowest-weight representations);
\item The principle unitary series representations $V_{\ui a,\epsilon}$ with weights $\{\epsilon+n | n \in \mathbb{Z} \}$, where $\epsilon \in \{ 0,\frac{1}{2} \}$ and $a\in \mathbb{R}$, $a>0$, on which the Casimir operator takes the value $-(\frac{1}{4} + a^2)$ (these representations are self-dual).
\end{enumerate}

The decomposition formula can now be stated precisely. Given any function $f(\hat{T})$ in the regular representation, define the coefficients
\begin{equation}
c_\mu^{m,n} = \int_{SU(1,1)} \ud \mu_T f(\hat{T})\left(t_\mu^{m,n}(\hat{T})\right)^\ast 
\end{equation}
where $\ud \mu_T$ is the Haar measure. Then 
\begin{multline}
f(\hat{T}) \propto \sum_{l\in \mathbb{N}/2} (l-\tfrac{1}{2}) \sum_{m,n\geq l} \left( c_{l,+}^{m,n} t_{l,+}^{m,n} (\hat{T}) + c_{l,-}^{-m,-n} t_{l,-}^{-m,-n} (\hat{T}) \right) + \\
\sum_{\epsilon \in \{ 0, 1/2 \}} \int_0^\infty \ud a \ a \tanh(\pi(a+\ui \epsilon)) \sum_{m,n \in \mathbb{Z}+\epsilon} c_{\ui a,\epsilon}^{m,n} t_{\ui a,\epsilon}^{m,n}(\hat{T})
\label{DecompGeneral}
\end{multline}
With appropriate choice of Haar measure we can set the constant of proportionality to one. The prefactor $(l-\tfrac{1}{2})$ in the first sum and the term $a \tanh(\pi(a+\ui \epsilon))$ in the integral are related to the normalization of the functions $t_\mu^{m,n}$. They form what is known as the \emph{Plancherel measure} of $SU(1,1)$.

For the delta function we have
\begin{equation}
c_\mu^{m,n} = \int_{SU(1,1)} \ud \mu_T \ \delta(\hat{T})\left(t_\mu^{m,n}(\hat{T})\right)^\ast = \left(t_\mu^{m,n}(\hat{1})\right)^\ast = \braket{n^\mu}{m^\mu} = \delta_{mn}
\end{equation}
and thus
\begin{multline}
\delta(\hat{T}) = \sum_{l\in \mathbb{N}/2} (l-\tfrac{1}{2}) \sum_{m\geq l} \left( t_{l,+}^{m,m} (\hat{T}) + t_{l,-}^{-m,-m} (\hat{T}) \right) + \\
\sum_{\epsilon \in \{ 0, 1/2 \}} \int_0^\infty \ud a \ a \tanh(\pi(a+\ui \epsilon)) \sum_{m \in \mathbb{Z}+\epsilon} t_{\ui a,\epsilon}^{m,m}(\hat{T})
\label{DecompDelta}
\end{multline}

Turning back to equation (\ref{ProbDiffusion}), the operator on the right-hand side can be written
\begin{equation}
\frac{1}{4\lambda} \big( 2{\hat{L}_z}^2 + {\hat{L}_x}^2 + {\hat{L}_y}^2 \big) = \frac{1}{\lambda}\left( \hat{c} - 3 \hat{L}_0^2 \right)
\end{equation}
and is therefore diagonized in the basis $t_\mu^{m,n}$. Hence the solution of equation (\ref{ProbDiffusion}) is given by
\begin{eqnarray}
P(\hat{T},x-x') & = & \exp\left[ \tfrac{1}{\lambda} (\hat{c} - 3 \hat{L}_0^2)(x-x') \right] \ket{\delta} \nonumber \\
& = & \sum_{l\in \mathbb{N}/2} (l-\tfrac{1}{2}) \sum_{m\geq l} \ue^{(l(l+1)-3m^2)(x-x')/\lambda} \left( t_{l,+}^{m,m} (\hat{T}) + t_{l,-}^{-m,-m} (\hat{T}) \right) + \nonumber \\
& & \sum_{\epsilon \in \{ 0, 1/2 \}} \int_0^\infty \ud a \ a \tanh(\pi(a+\ui \epsilon)) \sum_{m \in \mathbb{Z}+\epsilon} \ue^{-(1/4 + a^2 + 3m^2)(x-x')/\lambda} t_{\ui a,\epsilon}^{m,m}(\hat{T}) \nonumber \\ 
\label{TmatrixDist}
\end{eqnarray}
This is the final expression for the distribution of transfer matrices, exact in the limit of weak disorder, and forms the central result of this paper. In the following section we relate the functions $t_\mu^{m,m}(\hat{T})$ to more well known functions by choosing an explicit parametrization of the transfer matrices, and discuss some of the properties and limits of the distribution.

\section{Properties and limits of the probability distribution}\label{SecProperties}

\subsection{Form of the functions $t_\mu^{m,n}(\hat{T})$}

The parametrization of $SU(1,1)$ in terms of Euler angles is given by
\begin{eqnarray}
\hat{T}(\phi,\theta,\psi) & = & \ue^{\hat{L}_z \phi/2} \ue^{\hat{L}_x \theta/2} \ue^{\hat{L}_z \psi/2} \nonumber \\
& = & \begin{pmatrix} \cosh \tfrac{\theta}{2} \ue^{\ui(\phi+\psi)/2} & \sinh \tfrac{\theta}{2} \ue^{\ui(\phi-\psi)/2} \\[6pt] \sinh \tfrac{\theta}{2} \ue^{\ui(-\phi+\psi)/2} & \cosh \tfrac{\theta}{2} \ue^{-\ui(\phi+\psi)/2} \end{pmatrix}
\end{eqnarray}
where $0 \leq \phi < 2\pi$, $0\leq \theta < \infty$, $-2\pi \leq \psi < 2\pi$, and the parametrization is unique except at the identity. With the Haar measure $\ud \mu_T = \frac{1}{8\pi^2}\ud \phi\ud\psi \ud(\cosh\theta)$ the constant of proportionality in (\ref{DecompGeneral}) is unity.

The functions $t_{l,\pm}^{m,m}(\hat{T})$ can then be written as \cite{Vilenkin1991},
\begin{equation}\label{DiscreteReptmm}
t_{l,\pm}^{m,m}(\hat{T}) = \ue^{\ui m (\phi+\psi)} (\cosh \tfrac{\theta}{2})^{-2l} {}_2 F_1(l-m,l+m;1;\tanh^2 \tfrac{\theta}{2})
\end{equation}
where $\pm$ takes the value $+$ for $m\geq l$ and the value $-$ for $m \leq -l$. Note that the expansion of the hypergeometric function terminates in this case (it gives the Jacobi polynomial $P_{m-l}^{(0,2l-1)}(1-2\tanh^2 \tfrac{\theta}{2})$). Similarly the functions $t_{\ui a,\epsilon}^{m,m}(\hat{T})$ can be written as
\begin{equation}\label{ContinuousReptmm}
t_{\ui a,\epsilon}^{m,m}(\hat{T}) = \ue^{\ui m (\phi+\psi)} (\cosh \tfrac{\theta}{2})^{-1+\ui 2 a} {}_2 F_1(\tfrac{1}{2}-m-\ui a,\tfrac{1}{2}+m-\ui a;1;\tanh^2 \tfrac{\theta}{2})
\end{equation}
Equations (\ref{DiscreteReptmm}) and (\ref{ContinuousReptmm}) can be confirmed by writing $\hat{c}$ and $\hat{L}_0$ as differential operators in ($\phi, \psi,\theta$), and showing that the functions are Eigenfunctions of these operators with the required Eigenvalues.

\subsection{Reflection and transmission amplitudes}

In terms of the reflection and transmission amplitudes for a particle approaching the disordered region $(x',x)$ from the left, $r$ and $t$ respectively,  the transfer matrix is given by
\begin{equation}
\hat{T}(x,x') = \begin{pmatrix} 1/t^\ast & r/t \\ r^\ast/t^\ast & 1/t \end{pmatrix}
\end{equation}
Comparing this with the parametrization in terms of Euler angles above, and setting $\phi_\pm = \frac{1}{2}(\phi \pm \psi)$, we see that $t = \ue^{\ui\phi_+}/\cosh\frac{\theta}{2}$ and $r = \ue^{\ui\phi}\tanh \frac{\theta}{2}$.

\begin{figure} 
\begin{center}
\begin{tabular}{lll}
\begin{picture}(110,80)
\put(0,80){a)}
\put(10,0){\includegraphics[
  width=3.7cm]{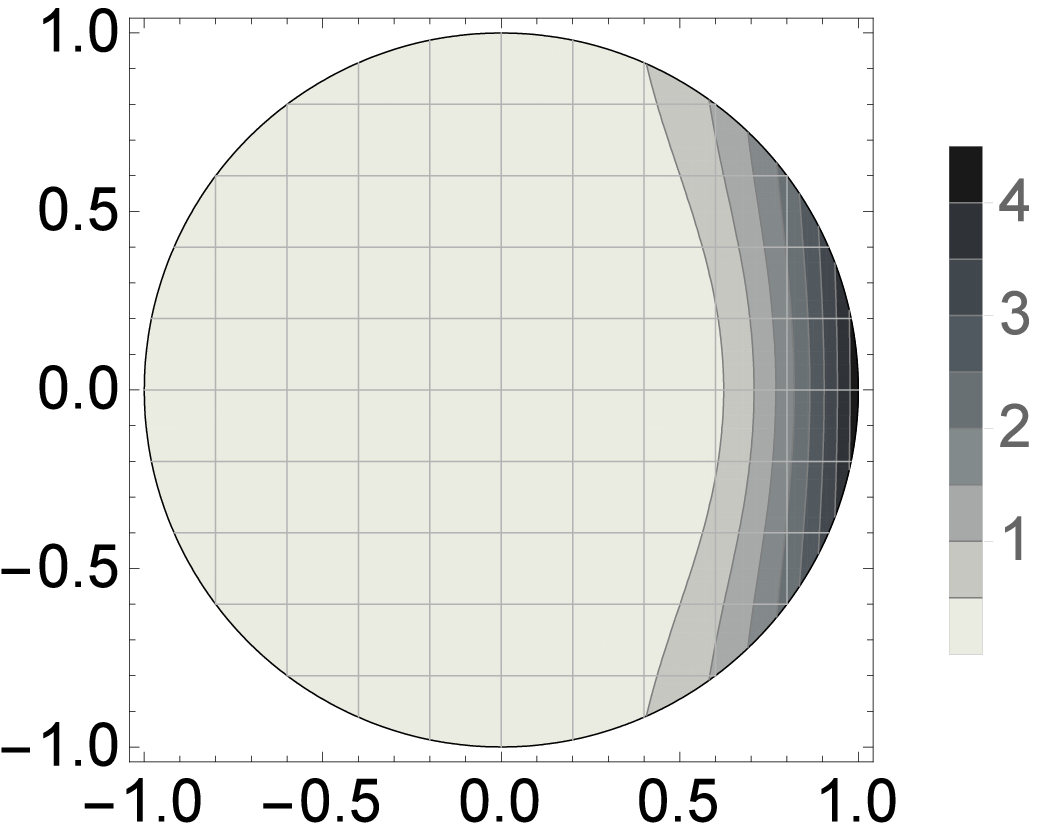} }
\end{picture} &
\begin{picture}(110,80)
\put(0,80){b)}
\put(10,0){\includegraphics[
  width=3.9cm]{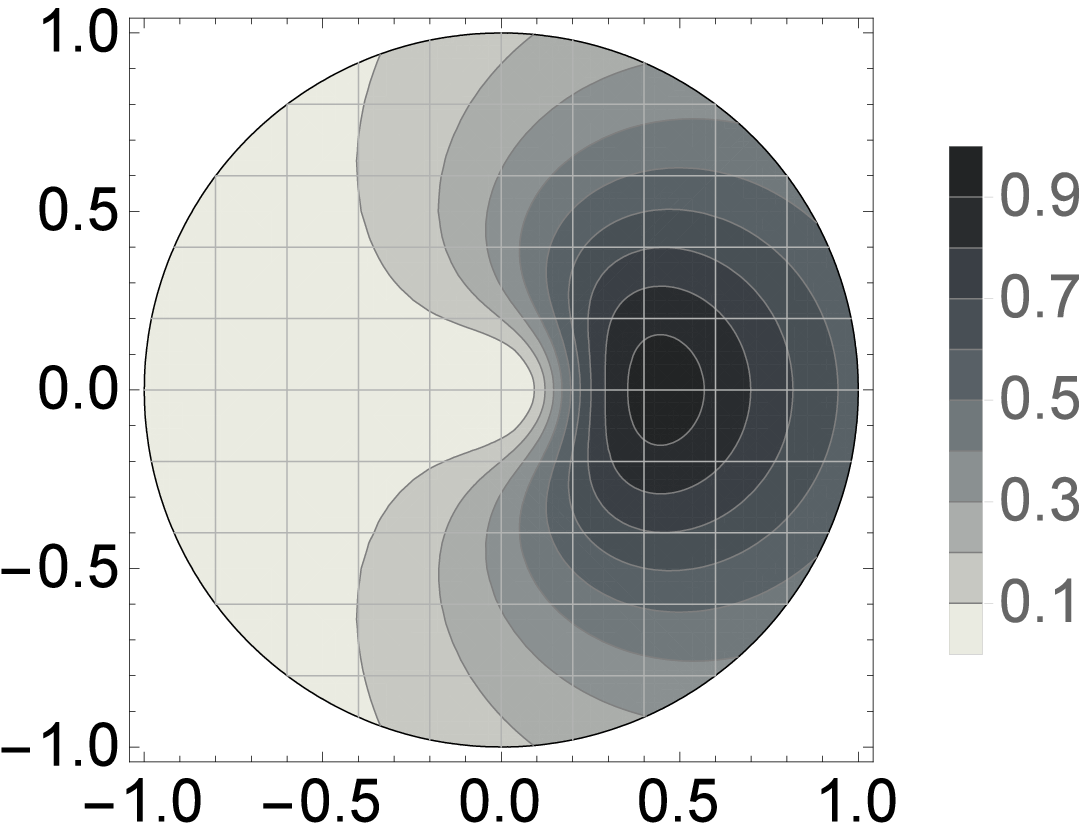} }
\end{picture} &
\begin{picture}(110,80)
\put(0,80){c)}
\put(10,0){\includegraphics[
  width=3.7cm]{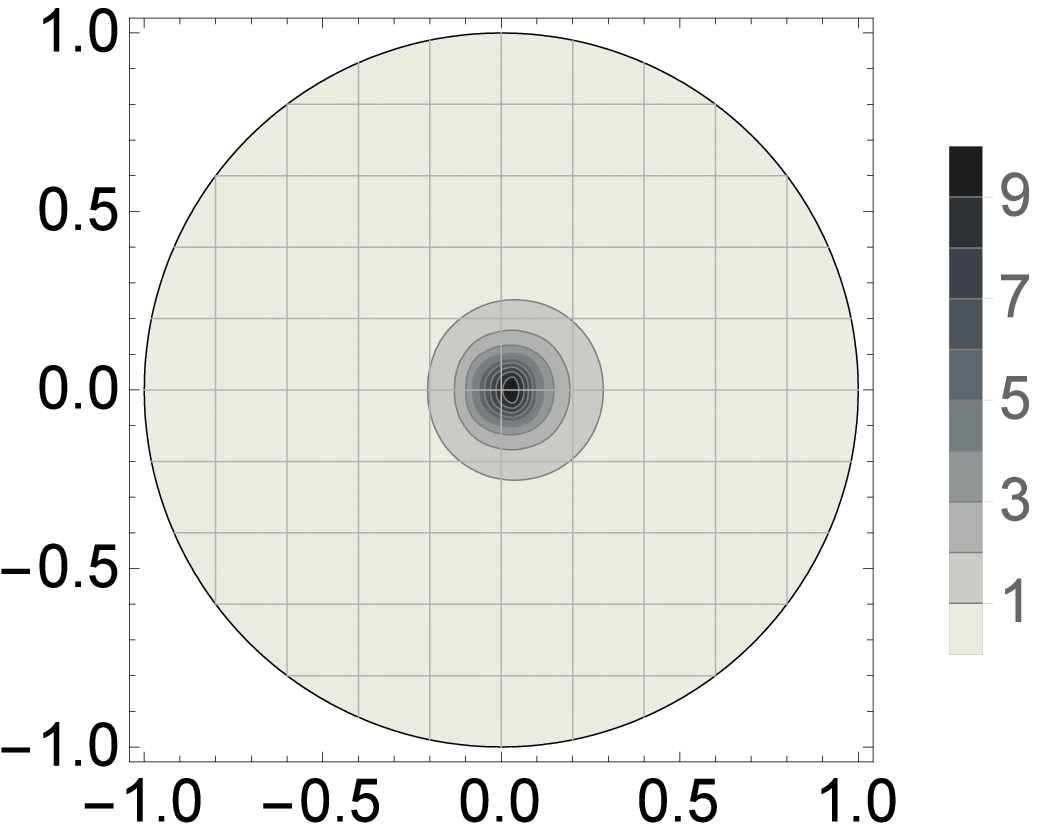} }
\end{picture}
\end{tabular}
\caption{Distribution of the transmission amplitude $t$, represented in the unit disc of the complex plane. Values of $L/\lambda$ are (a) 0.3, (b) 1.0, and (c) 3.0.}
\label{FiguretDist}
\end{center}
\end{figure}

The fact that the functions $t_\mu^{m,m}$ depend only on $\phi_+$ and $\theta$ implies that the probability of the transfer matrix is independent of the phase of $r$. Further, since $|r|^2 = 1- |t|^2$, the distribution can be expressed as a function of $t$ alone. Figure \ref{FiguretDist} shows some plots of the probability density as a function of $t$ for various ratios of $(x-x')/\lambda$.

\subsection{Distribution of the resistance}

The distribution of the resistance $\rho = |r|^2/|t|^2 = \sinh^2 \tfrac{\theta}{2}$ can be found by integrating (\ref{TmatrixDist}) over $\psi$ and $\phi$. Only the functions $t_{\ui a,0}^{0,0}$ contribute and the distribution is found to be
\begin{equation}\label{Resistivity1}
P(\theta,L) = \int_0^\infty \ud a \ a \tanh(\pi a) \ue^{-(1/4+a^2)L/\lambda} t_{\ui a,0}^{0,0}(\theta)
\end{equation}
(Note that we have kept the measure $\ud \cosh \theta = 2\ud \rho$). This expression can be simplified by choosing an integral representation for the hypergeometric function (derived in appendix \ref{app1}):
\begin{equation}
t_{\ui a,0}^{0,0}(\theta) = \frac{\cosh(\pi a)}{\sqrt{2} \pi} \int_0^\infty \frac{\cos (av)}{\sqrt{\cosh v + \cosh \theta}}  \ud v
\end{equation}

With this expression for the $t_{\ui a,0}^{0,0}$, the $a$-integral in (\ref{Resistivity1}) is now of Gaussian form. Evaluating it yields
\begin{equation}
P(\theta,L) = \frac{\ui}{4\sqrt{2\pi}} \left( \frac{\lambda}{L} \right)^{3/2} \ue^{-L/4\lambda} \int_{-\infty}^\infty \frac{(v-\ui\pi)\ue^{-(v-\ui\pi)^2\lambda/4L}}{\sqrt{\cosh v + \cosh \theta}} \ud v
\end{equation}
Shifting the integration contour by $+\ui\pi$ and taking proper account of the branch cut between $v=(-\theta-\ui\pi)$ and $v=(\theta -\ui\pi)$ gives
\begin{equation}\label{ResistDist}
P(\theta,L) = \frac{1}{4\sqrt{\pi}} \left( \frac{\lambda}{L} \right)^{3/2} \ue^{-L/4\lambda} \int_\theta^\infty \frac{v\ue^{-v^2\lambda/4L}}{\sqrt{\cosh^2 \frac{v}{2} - \cosh^2 \frac{\theta}{2}}} \ud v
\end{equation}

This expression is identical to the one first calculated by Abrikosov in \cite{Abrikosov1981}. There a Fokker-Planck equation was derived for the distribution of resistance as a function of length, an approach very similar in spirit to the one taken in section \ref{SecDiffusion}. Indeed, if we had chosen the representation of $L^2$-functions on the set of cosets $SU(1,1)/\langle\hat{L}_z\rangle$ instead of the regular representation, the two approaches would be exactly equivalent. (Here $\langle\hat{L}_z\rangle$ denotes the subgroup generated by $\hat{L}_z$, and the set of cosets thus produced can be parametrized by the reflection amplitude $r$, for details see \cite{MyThesis}.)

\subsection{The weak-localized limit}

In the weak-localized limit, $L\gg \lambda$, the contribution of states with large $m$ in (\ref{TmatrixDist}) is exponentially suppressed. Since $t^{m,m}\sim \cos (2m\phi_+)$, the distribution becomes approximately uniform in $\phi_+$. The contribution of states with $m=0$ has already been calculated in the previous section. The contribution of states with $m=\frac{1}{2}$ can be calculated in a similar way (see the appendix), giving
\begin{equation}\label{Phalf}
P^{(1/2)}(\theta, \phi_+,L) = \frac{1}{2\sqrt{\pi}} \left(\frac{\lambda}{L}\right)^{3/2} \ue^{-L/\lambda} \frac{\cos \phi_+}{\cosh\frac{\theta}{2}} \int_\theta^\infty \frac{v\cosh\frac{v}{2}\ue^{-v^2\lambda/4L}}{\sqrt{ \cosh^2\frac{v}{2}-\cosh^2\frac{\theta}{2}}} \ud v
\end{equation}
where $P^{(m)}$ stands for the total contribution of the $t^{m,m}$ states (i.e. the $2m$-th harmonic in $\phi_+$). Comparing this with (\ref{ResistDist}) we see that $P^{(1/2)}/P^{(0)} \sim \ue^{-3L/4\lambda}$ in the weak-localized limit. 

The higher harmonics can be found in a similar way, but the expressions become increasingly complicated owing to the appearance of poles in the integral representation for $t^{m,m}$ and the inclusion of the discrete-series representations. In particular, it is not easy to give an expression equivalent to (\ref{Phalf}) for general $m$.

As found in \cite{Abrikosov1981}, for $L \gg \lambda$ the distribution of $P^{(0)}$ becomes approximately Gaussian in $\theta$,
\begin{equation}
P^{(0)}(\theta,\phi_+,L) \approx \frac{1}{2\sqrt{\pi}} \left( \frac{\lambda}{L} \right)^{1/2} \ue^{-(\theta - L/\lambda)^2\lambda/4L}
\end{equation}
with respect to measure $\ud \theta$. We can show a similar result also holds for the higher harmonics. For $L \gg \lambda$ the integrand in (\ref{Phalf}) is only significant for $u = v-\theta \ll \theta$. Keeping only leading order terms in $u/\theta$ leads to
\begin{multline}
P^{(1/2)}(\theta,\phi_+,L) \approx \frac{1}{2\sqrt{2\pi}} \left( \frac{\lambda}{L} \right)^{3/2} \ue^{-3L/4\lambda}\ue^{-(\theta - L/\lambda)^2\lambda/4L} \cos\phi_+ \\ \int_0^\infty \frac{\theta \ue^{-(\theta \lambda /2L -1/2)u }}{\sqrt{\ue^u-1}}\ud u
\end{multline}
with respect to $\frac{1}{2\pi} \ud \phi_+ \ud \theta$. Making a change of variables $y=(\ue^u-1)^{1/2}$ gives
\begin{multline}
P^{(1/2)}(\theta,\phi_+,L) \approx \frac{1}{\sqrt{2\pi}} \left( \frac{\lambda}{L} \right)^{1/2} \ue^{-3L/4\lambda} \ue^{-(\theta - L/\lambda)^2\lambda/4L} \cos \phi_+ \\
\int_{0}^{\infty} (1+y^2)^{-\lambda \theta /2L -1/2} \ud y
\end{multline}
Setting $\theta = L/\lambda$ in the integrand allows the integral to be taken, leading to a Gaussian distribution in $\theta$,
\begin{equation}
P^{(1/2)}(\theta,\phi_+,L) \approx \frac{\sqrt{\pi}}{2} \left( \frac{\lambda}{L} \right)^{1/2} \ue^{-3L/4\lambda} \ue^{-(\theta - L/\lambda)^2\lambda/4L} \cos \phi_+
\end{equation}
The mean and variance of the Gaussian distribution are the same as for $P^{(0)}$.

\subsection{The far-ballistic limit}

For $\lambda \gg L$, the distribution is sharply peaked around $\theta = \phi_+ = 0$. The exponential factors in (\ref{TmatrixDist}) decay slowly, and we can replace the sums on $m$ and $l$ with integrals, and assume $\alpha,l,m \gg 1$. Specifically, approximating
\begin{align}
t_l^{m,m} &\approx \ue^{\ui 2 m \phi_+} {}_0F_1(1;(l^2-m^2)\tanh^2 \tfrac{\theta}{2}) \\
t_{\alpha,\epsilon}^{m,m} &\approx \ue^{\ui 2 m \phi_+} {}_0F_1(1;-(\alpha^2+m^2)\tanh^2 \tfrac{\theta}{2}) \end{align}
and
\begin{multline}
\sum_{l\in \mathbb{N}/2} (l-\tfrac{1}{2}) \sum_{m\geq l} \ue^{(l(l+1)-3m^2)L/\lambda} \cdots \to \\ 2\int_0^\infty \ud l \ l \ue^{l^2 L/\lambda} \int_l^\infty \ud m \  \ue^{-3m^2L/\lambda} \cdots 
\end{multline}
\begin{multline}
\sum_{\epsilon \in \{ 0, 1/2 \}} \int_0^\infty \ud a \ a \tanh(\pi(a+\ui \epsilon)) \sum_{m \in \mathbb{Z}+\epsilon} \ue^{-(1/4 + a^2 + 3m^2)L/\lambda} \cdots \to \\
2 \int_0^\infty \ud \alpha \ \alpha \ue^{-\alpha^2 L/\lambda} \int_{-\infty}^{\infty}  \ud m \ \ue^{-3m^2L/\lambda} \cdots
\end{multline}
leads to integrals which can be evaluated exactly. The result is
\begin{equation}\label{ProbDistBall}
P(\theta,\phi_+,L) \approx \sqrt{\frac{\pi}{2}} \left( \frac{\lambda}{L} \right)^{3/2} \theta \, \ue ^{-\frac{\lambda}{4L}(\theta^2+2\phi_+^2)}
\end{equation}
with respect to measure $\frac{1}{2\pi} \ud \phi_+ \ud \theta$.

\section{Persistent current in a flux-threaded ring}\label{SecRing}

Given that the distribution of the transmission coefficient is a previously known result, the additional information contained in the main result of this paper, (\ref{TmatrixDist}), is the dependence of the probability on the phase of $t$. In an open, simply-connected wire, it turns out that this phase is usually irrelevant in the calculation of physical quantities. For example, the absolute value of the Green's function $G_E(x,x')$ is independent of the phase of the transmission amplitude, and therefore the density-density correlators are likewise independent. Similarly, the distribution of the local density of states, $\rho(x) \sim \text{Im}(G_E(x,x))$ can be found directly from the distribution of the transmission coefficient (as has been explicitly demonstrated in \cite{Schomerus(Beenakker)2002}).

For an application where the full distribution of $t$ is necessary, we therefore turn to a closed, disordered ring. An interesting feature of such rings is the appearance of a persistent current when threaded by a magnetic flux \cite{Buttiker1983}. More recent theoretical interest in such systems \cite{Ginossar2010,Bary-Soroker(Imry)2010,Danon(Brouwer)2010} has been spurred by advances in experimental techniques \cite{Bleszynski-Jayich2009,Castellanos-Beltran2013} which allow the measurement of persistent current in a single mesoscopic ring. 

Below, we calculate the average contribution to persistent current from states with energy between $E$ and $E+\ud E$. The contribution can be written $i(E)\ud E$ where
\begin{equation}\label{CurrentS1}
i(E) = \sum_n \delta(E-E_n) j_n = \sum_n \delta(E-E_n) \pd{E_n}{\Phi}
\end{equation}
Here $n$ denotes a sum over single-electron states, $j_n$ is the current of the $n$-th state, and $\Phi$ denotes the total flux through the ring \cite{Riedel1989}.

\begin{figure}
\begin{center}
\begin{tabular}{lll}
\begin{picture}(110,80)
\put(0,80){a)}
\put(5,0){\includegraphics[
  width=4cm]{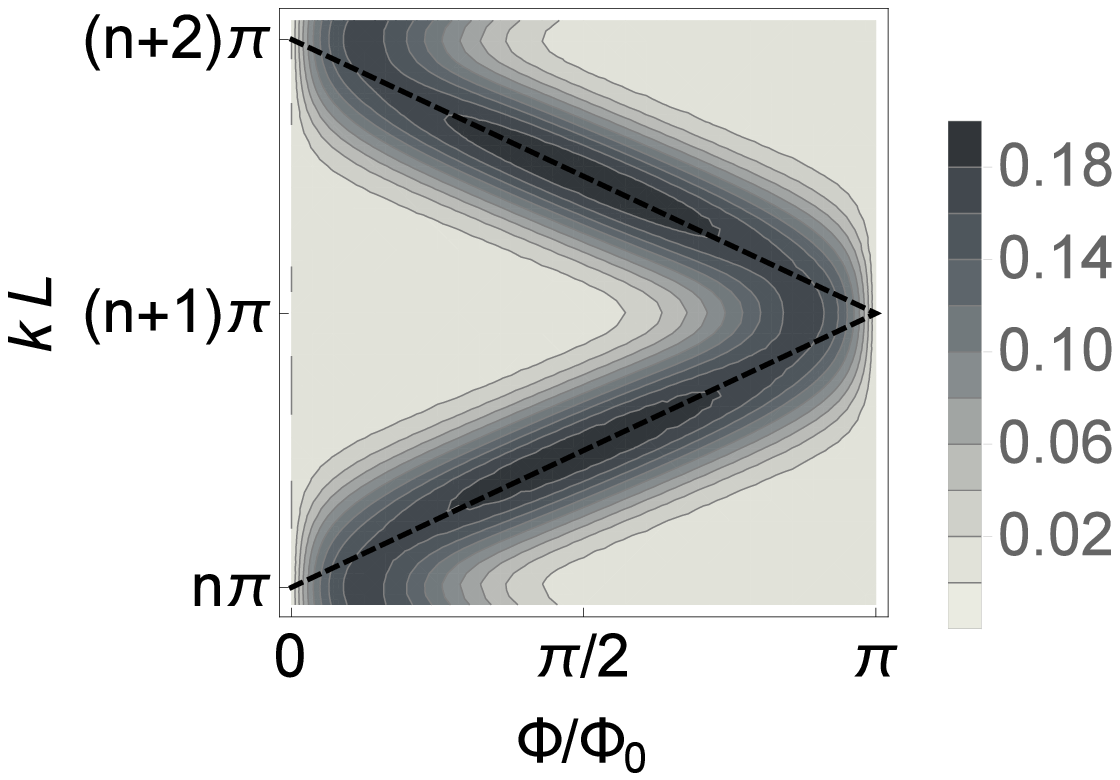} }
\end{picture} &
\begin{picture}(110,80)
\put(0,80){b)}
\put(5,0){\includegraphics[
  width=4cm]{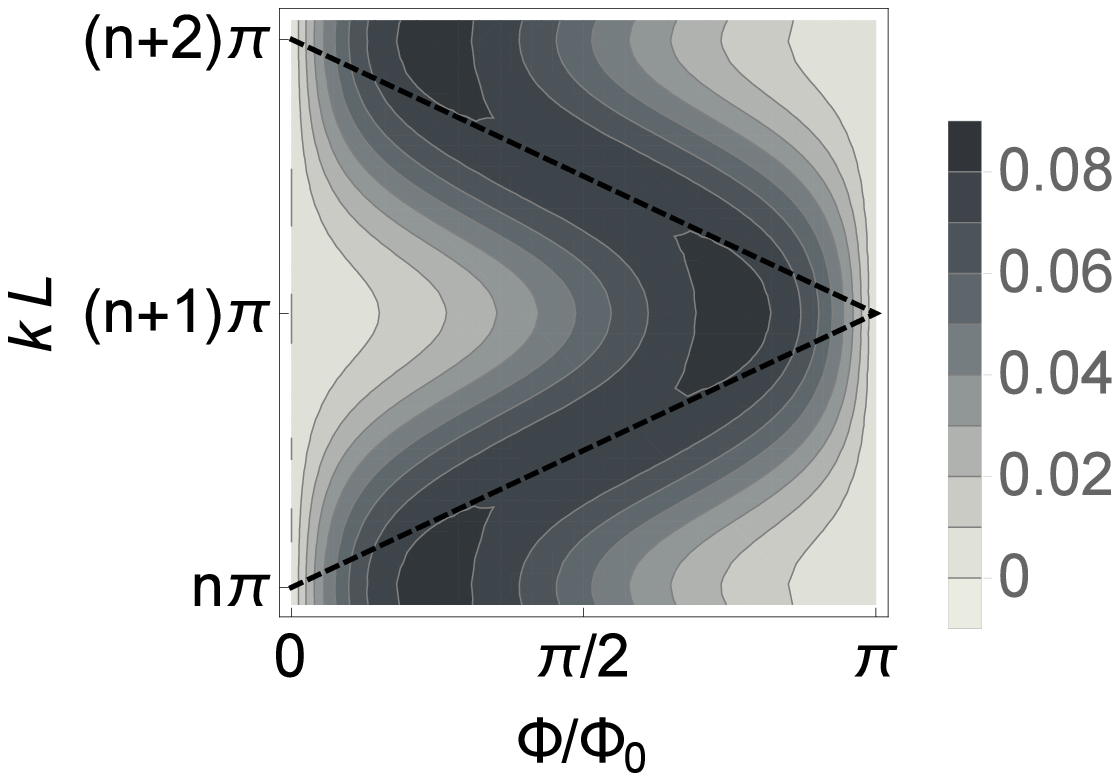} }
\end{picture} &
\begin{picture}(110,80)
\put(0,80){c)}
\put(5,0){\includegraphics[
  width=4cm]{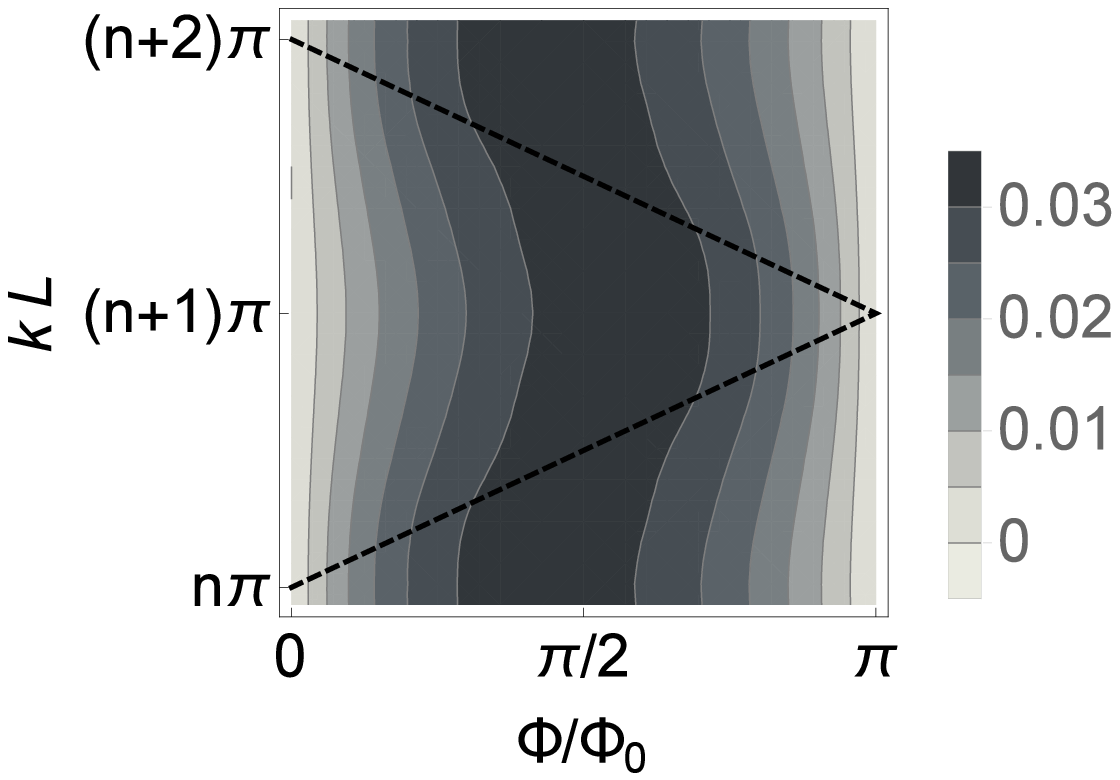} }
\end{picture} \\
\begin{picture}(110,80)
\put(0,70){d)}
\put(5,0){\includegraphics[
  width=3.5cm]{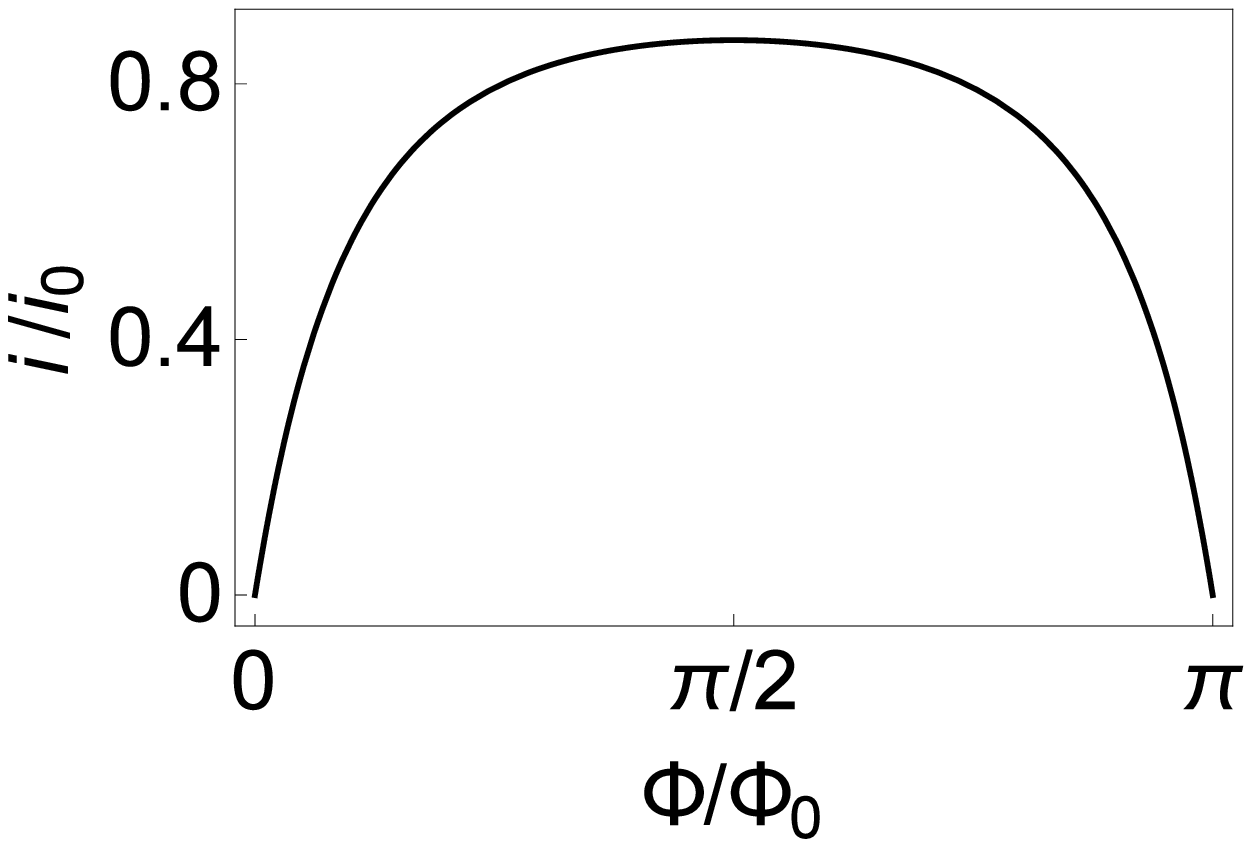} }
\end{picture} &
\begin{picture}(110,80)
\put(0,70){e)}
\put(5,0){\includegraphics[
  width=3.5cm]{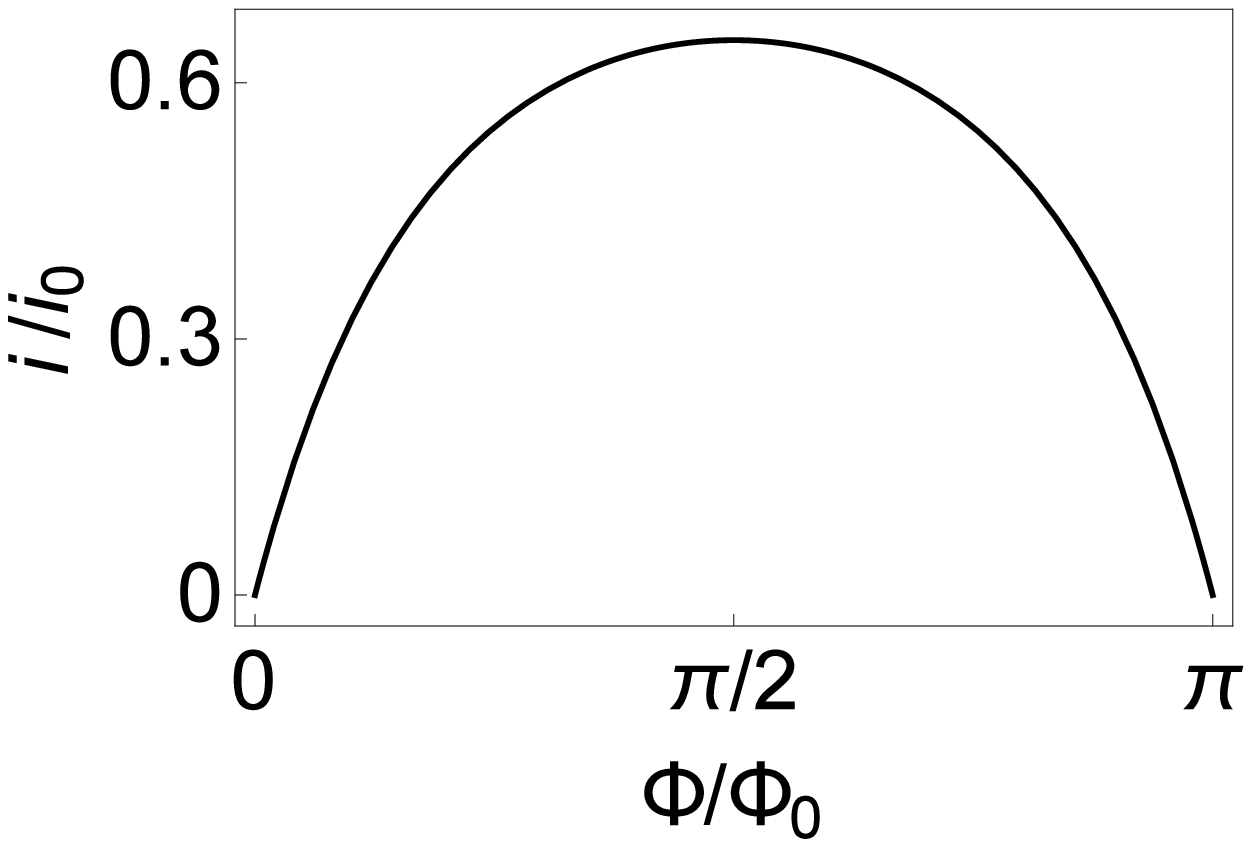} }
\end{picture} &
\begin{picture}(110,80)
\put(0,70){f)}
\put(5,0){\includegraphics[
  width=3.5cm]{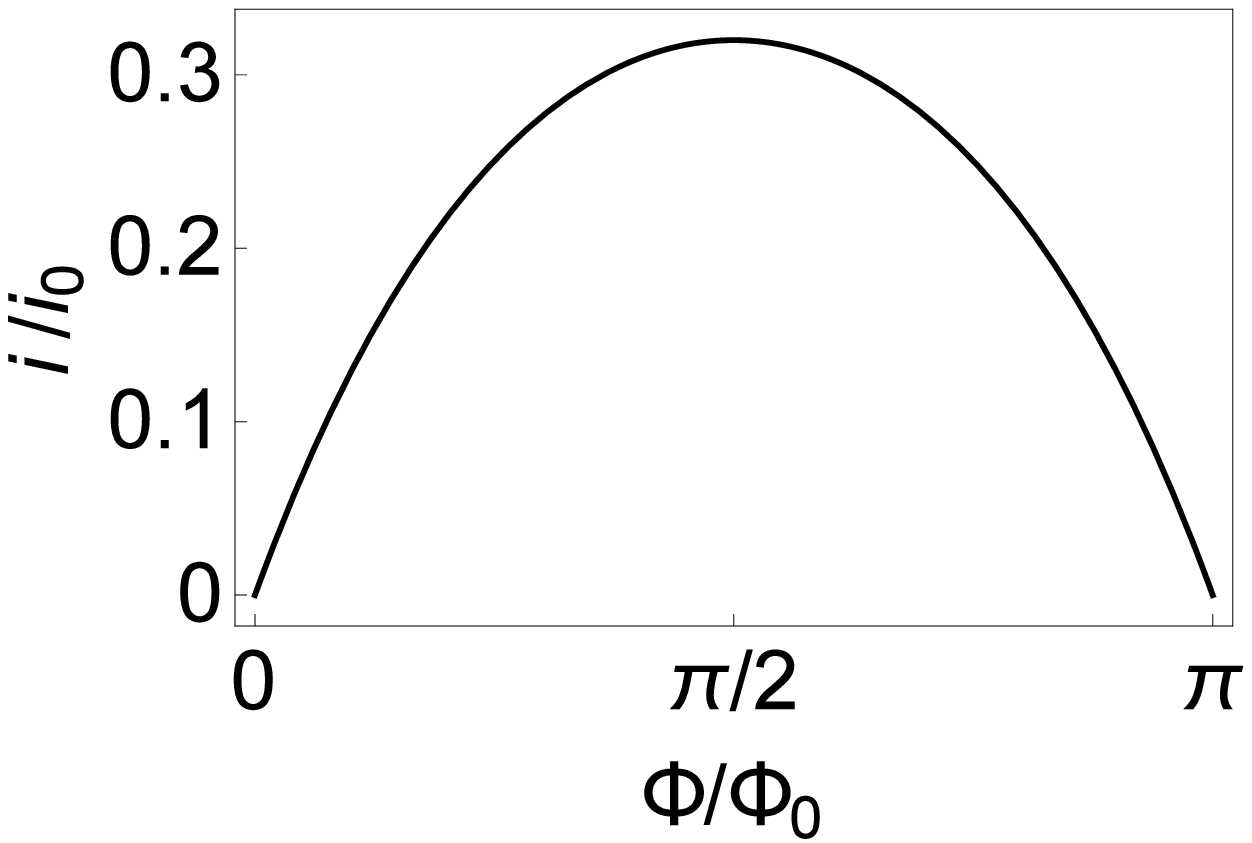} }
\end{picture}
\end{tabular}
\caption{a)--c): Plots of $i(E)$ as a function of magnetic flux and energy $E=k^2$ for $L/\lambda = 0.3,1.0,3.0$, respectively. The dashed line shows the positions of the energy levels of the clean system. d)--f): the average current per level as a fraction of the single-level current of the clean system. Values of $L/\lambda$ are as in a)--c).}
\label{FigureCurrent}
\end{center}
\end{figure}

The presence of the magnetic flux imposes twisted boundary conditions on the wavefunction: $\psi_n(L) = \ue^{\ui 2\pi \Phi/\Phi_0} \psi_n(0)$, $\psi_n'(L) = \ue^{\ui 2\pi \Phi/\Phi_0} \psi_n'(0)$, where $\Phi_0 = h/e$ is the magnetic flux quantum. This implies that $\left( \begin{smallmatrix} \phi_R(0) \\ \phi_L(0) \end{smallmatrix} \right)_n$ is an Eigenfunction of $\ue^{k_n L \hat{L}_z}\hat{T}(L,0)$ with Eigenvalue $\ue^{\ui 2\pi \Phi/\Phi_0}$. Parametrizing the transfer matrix by Euler angles, the Eigenvalue equation takes the form
\begin{equation}\label{CurrentS2}
\cos \left( 2\pi \tfrac{\Phi}{\Phi_0}\right) = \cosh \tfrac{\theta}{2} \cos(\phi_+ +k_n L)
\end{equation}
In what follows, for notational convenience we choose units $e=1$ and set $2\pi \frac{\Phi}{\Phi_0} = KL$. Differentiating (\ref{CurrentS2}) with respect to $\Phi$ and substituting the result into (\ref{CurrentS1}) gives,
\begin{equation}\label{CurrentS3}
|i(E)| =|\sin KL| \ \delta \left( \cosh \tfrac{\theta}{2} \cos(\phi_+ +k L) - \cos KL \right)
\end{equation}
as found in \cite{Kolokolov1994}. Averaging over $P(\theta,\phi_+,L)$, we find
\begin{multline}
\expected{|i(E)|} = \frac{1}{2\pi} \int_1^\infty \ud (\cosh \theta) \frac{\sin KL}{\sqrt{\cosh^2 \tfrac{\theta}{2} - \cos^2 KL}} \times \\ \Big[ P\left(\theta , \cos^{-1} \big( \tfrac{\cos KL}{\cosh \theta/2}\big) - kL,L \right) 
+ P\left(\theta , -\cos^{-1} \big( \tfrac{\cos KL}{\cosh \theta/2} \big) -kL ,L \right) \Big]
\label{Current1}
\end{multline}
Figure \ref{FigureCurrent} shows the average persistent current as a function of $E$ for different values of $L/\lambda$.

In the weak-localized and far-ballistic limits it is possible to simplify the above expression. In the far-ballistic limit, setting $\cosh\tfrac{\theta}{2} = \frac{\cos KL}{\cos u}$ and using (\ref{ProbDistBall}) we find
\begin{multline}
\expected{|i(E)|} \approx \sqrt{\frac{2}{\pi}} \left( \frac{\lambda}{L}\right)^{3/2} \sin KL \int_{KL}^{\pi/2} \frac{\ud u}{\cos u}\ue^{-(1-cos^2 u / \cos^2 KL)\lambda/L} \times \\
\left( \ue^{-(u^2 -\tilde{k}L)^2\lambda/2L} + \ue^{-(u^2 +\tilde{k}L)^2\lambda/2L} \right)
\end{multline}
Here we have assumed $0 \leq KL \leq \pi$ (this is no restriction since the persistent current is an odd function of $KL$), and set $\tilde{k}L/2\pi$ as the fractional part of $kL/2\pi$. The integrand is only significant when $u \approx KL \approx \pm \tilde{k}L$ or when $u \approx \frac{\pi}{2}$. The pole at $\frac{\pi}{2}$ is a consequence of the use of the approximate formula $\theta^2 \approx 4 \tanh^2 \frac{\theta}{2}$, and should be ignored. Setting $\tilde{k} = \pm K +\delta k$, $u = KL + \eta$ and keeping only leading terms in $\delta k$ and $\eta$, for $KL \gg \sqrt{\frac{L}{\lambda}}$ and $\pi - KL \gg \sqrt{\frac{L}{\lambda}}$ we obtain
\begin{align}
\expected{|i(E)|} & \approx \sqrt{\frac{2}{\pi}}  \left( \frac{\lambda}{L}\right)^{3/2} \ue^{-(\delta k L)^2 \lambda/2L}  \tan KL \int_0^\infty \ue^{-\frac{2\lambda}{L}\eta\tan KL} \ud \eta \nonumber \\
& = \frac{1}{\sqrt{2\pi}} \left( \frac{\lambda}{L}\right)^{1/2} \ue^{-(\delta k L)^2 \lambda/2L}
\end{align}
In the absence of disorder, the energy levels are given by $\tilde{k} = \pm K$ with current $2\frac{|k|}{L}$. Thus in the far-ballistic limit the average current per level is unchanged, and the energy-levels are Gaussian distributed around those of a clean wire (see figure \ref{FigureCurrent}).

The situation is different if $KL \lesssim \sqrt{\frac{L}{\lambda}}$ or $\pi - KL \lesssim \sqrt{\frac{L}{\lambda}}$, as should be expected since $KL = 0 ,\pi$ mark the positions of the level crossings in the clean system. For $KL \lesssim \sqrt{\frac{L}{\lambda}}$, approximating $\frac{\cos^2 u}{\cos^2 KL} \approx (KL)^2-u^2$ we obtain
\begin{multline}
\expected{|i(E)|} = \frac{1}{\sqrt{3}} K \lambda\ue^{((3(KL)^2- \tilde{k}L)^2)\lambda/3L} \times \\
\left[ 1 - \erf{(K-\tfrac{\tilde{k}}{3})\sqrt{3\lambda L}} - \erf{(K+\tfrac{\tilde{k}}{3})\sqrt{3\lambda L}} \right] 
\end{multline}
where $\erf{x}$ denotes the error function.

In the weak localized limit, substituting equation (\ref{ResistDist}) for $P(\theta,\phi_+,L)$ into equation (\ref{Current1}) gives
\begin{multline}
\expected{|i(E)|} = \frac{1}{2} \sin KL \left( \frac{\lambda}{\pi L} \right)^{3/2} \ue^{-L/4\lambda} \\
\int_1^\infty \ud(\cosh\theta) \int_{\theta}^{\infty} \ud v \frac{v \ue^{-v^2\lambda/4L}}{\sqrt{\cosh\theta - \cos 2KL}\sqrt{\cosh v - \cosh \theta}}
\end{multline}
exchanging the order of integration allows the $\theta$-integral to be taken. An integration by parts in $v$ then leads to
\begin{equation}\label{CurrentWeakLoc}
\expected{|i(E)|} = \frac{1}{2\pi} \sin^2 KL \left( \frac{\lambda}{\pi L} \right)^{1/2} \ue^{-L/4\lambda} \int_0^\infty \ud v \frac{\cosh \frac{v}{2} \ue^{-v^2\lambda/4L}}{\cosh^2 \frac{v}{2} - \cos^2 KL}
\end{equation}
which agrees with the result found in \cite{Kolokolov1994}. 

Notice that the transmission phase $\phi_+$ in equation (\ref{CurrentS3}) occurs only in the term $\cos (\phi_+ + kL)$, which is a fast oscillating function of $L$. Therefore variation in $L$ of just a single wavelength in the ensemble is enough to destroy the dependence of $\expected{|i(E)|}$ on $\phi_+$, in which case (\ref{CurrentWeakLoc}) is obtained for all strengths of disorder. 

Unfortunately the distribution of the transfer matrix is not enough to derive the distribution of $i(E)$. This can be seen from equation (\ref{CurrentS1}), as higher moments depend in a non-trivial way on $\pd{E_n}{\Phi}$ -- i.e. on the correlations between transfer matrices at nearby energies. It is possible to obtain the distribution of $i(E)$ in the case of a ring connected to an electron reservoir (e.g. as modelled in \cite{Buttiker1985}), but the result is sensitive to the nature of the coupling. Connecting the ring to a reservoir also allows the distribution of local density of states to be obtained (some results for the local density of states in the closed system have been found in \cite{Feldmann2000}).

\section{Discussion}

This paper has introduced a new method for obtaining the transport statistics of disordered one-dimensional systems. The approach can be summarized as follows: firstly identify the symmetry group of the transfer matrix (here $SU(1,1)$). Secondly relate the disordered potential to a diffusion equation on this group (equation (\ref{TavEvolution})). Thirdly choose a representation on which the group acts (here the regular representation). Finally, solve the diffusion equation on that representation using the representation theory (in this case we were able to exactly diagonalize the differential operator by expressing it in terms of the Casimir invariant of the group). In principle the same approach can be applied to any system which admits a description in terms of a transfer matrix. The simplest generalization would be to $n$-channel wires, where the $SU(1,1)$ group is replaced by $SU(n,n)$ (or the subgroup $Sp(2n,\mathbb{R})$ in systems with time-reversal symmetry).

The representation used can be tailored to the calculation in question. For example, as shown in \cite{Me2005}, the $n$-th moment of the reflection coefficient can be found in terms of the group action on the (finite-dimensional) spin-$n$ representation. As mentioned in the text, the full distribution of the transmission coefficient can likewise be found by considering the space of $L^2$-functions on the set of cosets $SU(1,1)/\langle \hat{L}_z \rangle$. Generalizing to the $n$-channel case, the group action on the set of cosets $U(n,n)/(U(n)\otimes U(n))$ yields an equation for the distribution function of the $n$ transmission coefficients, in a manner analogous to the DMPK technique \cite{Dorokhov1982,MPK1988} (here $U(n) \otimes U(n)$ is the block-diagonal subgroup generated by scattering between co-directional channels). The connection between the DMPK technique and harmonic analysis on coset spaces has previously been made \cite{Huffmann1990}, but the approach here is more general, as the evolution operator need not be proportional to the Laplace-Beltrami operator on a coset space; indeed the representation used need not be realizable as a representation of functions on a coset space at all.

Finally, given the general relationship between irreducible unitary representations of a Lie group and its co-adjoint orbits, and the existence of a natural symplectic form on these orbits (the Kirillov-Kostant-Souriau form \cite{Kirillov2004}), the approach should be amenable to semi-classical approximations. Given that exact solutions are unlikely in systems more complex than the one-dimensional wire considered here, the availability of approximation techniques is significant.

\section{Acknowledgements}

The theoretical basis of this work (including derivation of the distribution (\ref{TmatrixDist})) was developed as part of the author's PhD thesis under the supervision of Boris Muzykantskii.

This research was supported by the Catholic University of Korea research fund 2014.

\appendix

\section{Integral representations of the Hypergeometric functions}\label{app1}

The hypergeometric function ${}_2F_1(a,b;1;z)$ for $0\leq z < 1$ can be written
\begin{equation}
{}_2F_1(a,b;1;z) = \frac{1}{2\pi \ui} \oint w^{-1}(1-z/w)^{-a}(1-w)^{-b}\ud w
\end{equation}
where the integral is counter-clockwise around the unit circle in the complex plane, as can be checked by a Taylor expansion of the integrand. For $0\leq \text{Re}(a)<1$ the integral can be collapsed onto the branch cut along the real line from $0$ to $z$ to yield
\begin{equation}
{}_2 F_1(a,b;1;z) = \frac{\sin(\pi a)}{\pi} \int_0^z w^{-1} (z/w-1)^{-a}(1-w)^{-b} \ud w
\end{equation}
Inserting this into the expression for $t_{\ui a,0}^{0,0}$ gives
\begin{equation}
t_{\ui a,0}^{0,0}(\rho) = \frac{\cosh(\pi a)}{\pi}(\cosh \tfrac{\theta}{2})^{-1+\ui 2 a} \int_0^{\tanh^2 \frac{\theta}{2}} w^{-1} \left( (\tanh^2\tfrac{\theta}{2}-w) (1/w-1)\right)^{-1/2+\ui a} \ud w
\end{equation}
and making a change of variables to $v = \ln[(\sinh^2\tfrac{\theta}{2}-w\cosh^2\tfrac{\theta}{2}) (1/w-1)]$, we arrive at
\begin{equation}
t_{\ui a,0}^{0,0}(\rho) = \frac{\sqrt{2}\cosh(\pi a)}{\pi} \int_0^\infty \frac{\cos (av)}{\sqrt{\cosh v + \cosh \theta}}  \ud v
\end{equation}

A similar set of transformations for $t_{\ui a,\epsilon}^{\frac{1}{2},\frac{1}{2}}$ leads to the expression
\begin{equation}
t_{\ui a,0}^{\frac{1}{2},\frac{1}{2}}(\rho) = \frac{\sqrt{2}\sinh(\pi a)}{\pi \cosh\frac{\theta}{2}} \int_0^\infty \frac{\sinh\frac{v}{2}\ \sin (av)}{\sqrt{\cosh v + \cosh \theta}}  \ud v
\end{equation}

\input{main.bbl}

\end{document}

%% file: header.tex
\usepackage{graphicx}
\usepackage{amsmath}
\usepackage{amssymb}

\newcommand{\ud}{\text{d}}
\newcommand{\ue}{\text{e}}
\newcommand{\ui}{\text{i}}

\newcommand{\sd}[2]{\frac{\ud #1}{\ud #2}}

\newcommand{\pd}[2]{\frac{\partial #1}{\partial #2}}
\newcommand{\pdd}[3]{\frac{\partial^2 #1}{\partial #2 #3}}

\newcommand{\erf}[1]{\text{erf}\left( #1 \right)}

\newcommand{\ket}[1]{\left|  #1  \right\rangle}
\newcommand{\bra}[1]{\left\langle  #1  \right|}
\newcommand{\braket}[2]{\left\langle \left.  #1 \right| #2 \right\rangle}

\newcommand{\expected}[1]{\left\langle #1 \right\rangle}